\documentclass[prb,superscriptaddress,twocolumn,showpacs,preprintnumbers,amsmath,floatfix]{revtex4}
\usepackage{amssymb}
\usepackage{graphicx}

\begin{document}

\title{Gate-induced Gap in Bilayer Graphene Suppressed by Coulomb Repulsion}
\author{Jin-Rong Xu}
\affiliation{Shanghai Key Laboratory of Special Artificial Microstructure Materials and Technology, \\
School of Physics Science and engineering, Tongji University, Shanghai 200092, P.R. China}
\affiliation{School of Mathematics and Physics, Anhui Jianzhu University, Hefei, Anhui, 230601, P.R. China}
\author{Ze-Yi Song}
\affiliation{Shanghai Key Laboratory of Special Artificial Microstructure Materials and Technology, \\
School of Physics Science and engineering, Tongji University, Shanghai 200092, P.R. China}
\author{Hai-Qing Lin}
\affiliation{Beijing Computational Science Research Center, Beijing 100084, P.R. China}
\author{Yu-Zhong Zhang}
\email[Corresponding author. ]{Email: yzzhang@tongji.edu.cn}
\affiliation{Shanghai Key Laboratory of Special Artificial Microstructure Materials and Technology, \\
School of Physics Science and engineering, Tongji University, Shanghai 200092, P.R. China}
\affiliation{Beijing Computational Science Research Center, Beijing 100084, P.R. China}
\date{\today }

\begin{abstract}
\centerline{Abstract}
We investigate the effect of on-site Coulomb repulsion $U$ on the band gap of the electrically gated bilayer graphene by employing coherent potential approximation in the paramagnetic state, based on an ionic two-layer Hubbard model. We find that, while either the on-site Coulomb repulsion $U$ or the external perpendicular electric field $E$ alone will favor a gapped state in the bilayer graphene, competition between them will surprisingly lead to a suppression of the gap amplitude. Our results can be applied to understand the discrepancies of gap size reported from optical and transport measurements, as well as the puzzling features observed in angular resolved photoemission spectroscopic study.
\end{abstract}


\maketitle

\section{Introduction\label{Introduction}}

Bilayer graphene shares many of the interesting properties of monolayer graphene~\cite{CastroGeim,McCannKoshino}, but provides even richer physics due to the presence of massive chiral quasiparticles~\cite{Novoselov2006} and has even wider applications due to the possibility of controlling an infrared gap through doping and gating~\cite{Ohta2006,Oostinga,Castro2007,McCann,Min2007}. Although the band gap has now been observed in a number of different experiments~\cite{Ohta2006,Oostinga,Castro2007,Kuzmenko2009,Mak,Y.Zhang,Taychatanapat}, opening a possible way towards realizing graphene-based nanoelectronic and nanophotonic devices~\cite{Novoselov2012}, long-standing puzzle regarding the gap amplitude has not been solved between optical and transport measurements, i.e., while the gap observed in optics is up to 250~meV~\cite{Mak,Y.Zhang}, that derived from transport measurements is down to below 10~meV~\cite{Oostinga,Taychatanapat} upon applied electric field perpendicular to the graphene plane.

Therefore, tremendous effort has been made in understanding the discrepancy. Possible existences of mid-gap electronic states generated by material imperfections, such as vacancy~\cite{Castro2010}, disorder~\cite{Rossi2011,Miyazaki}, or structural distortions~\cite{Santos,Kim,Park2015}, edge states~\cite{Castro2008,Li2011}, were extensively discussed. These hypotheses are mainly based on a consensus that the low-energy behavior of electrons in bilayer graphene is well described by the tight-binding model without electronic interaction~\cite{McCannKoshino}.

But in fact, bilayer graphene has already been predicted to be unstable to the electronic interaction at half-filling due to a non-vanishing density of state present at the Fermi level~\cite{CastroGeim}. And the importance of electronic interaction has also been widely noticed experimentally in the bilayer graphene even in the absence of electric field~\cite{Maher,Martin,Mayorov,Velasco}. Though a number of theoretical studies have been made in searching the ground state with various symmetry breakings~\cite{Nandkishore2010,Zhang2012,Kharitonov2012,Lang2012,Tao2014,Sun2014}, to our knowledge, it is still lack of a theoretical study concerning the intrinsic and unavoidable correlation effects induced by electronic interaction~\cite{Wehling2011,Schueler2013} on the gate-induced gap.

Since the inconsistency of gap size between optical and transport measurements occurs irrespective of whether there is interaction-driven symmetry breaking, in this letter, we will investigate the effects of many-body correlation on the band gap by means of coherent potential approximation (CPA)~\cite{Elliot1974,Jarrell2001} in the paramagnetic state of gated Bernal stacked bilayer graphene which can be qualitatively described by an ionic two-layer Hubbard model. We will first present a phase diagram of the model where a novel interaction-driven metallic state appears between band and Mott insulating states. Then we find that the gated bilayer graphene is located in the vicinity of the phase boundary between the band insulator and the correlated metal.
The inconsistency of the gap size between optical and transport measurements~\cite{McCannKoshino,Mak,Y.Zhang,Oostinga,Taychatanapat,Ulstrup} can be resolved after many-body correlation is taken into account. Furthermore, our calculated spectrum reveals that exotic spectra observed in angular resolved photoemission spectroscopic study~\cite{Ohta2006,Kim} can also be simply attributed to the many-body effect without the need to assume lattice imperfection. Our study strongly suggests that many-body correlation should not be neglected in graphene-based systems.

Our paper is organized as follows. In Sec.~\ref{MM}, we describe the ionic two-layer Hubbard model and CPA method. In Sec.~\ref{RD}, we present our results, including phase diagram, density of states, self-energies, optical conductivities, and spectral functions. The relevance of our findings to various experimental observations is also discussed in this section. Finally, we do a summary in Sec.~\ref{Summary}.

\section{Method and model\label{MM}}

\begin{figure}[htbp]
\includegraphics[width=0.48\textwidth]{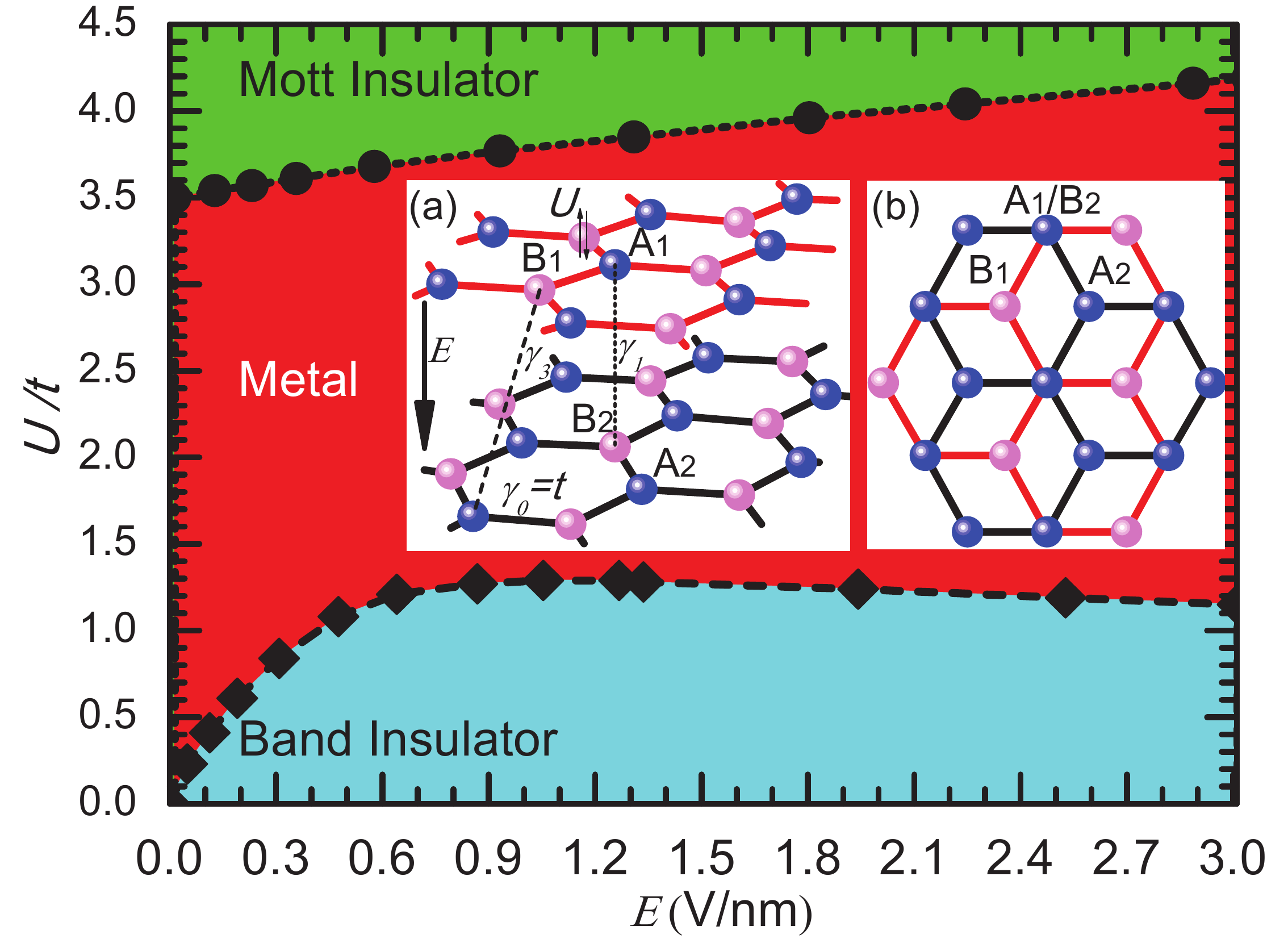}
\caption{(Color online) Phase diagram in the $U/t-E$ plane.
$\gamma_0 = t =2.7$~eV, $\gamma_1=0.4$~eV, and $\gamma_3=0.3$~eV
are fixed. An interaction-driven metallic state is sandwiched
between two gapped states. Inset (a) is the cartoon for side view
of bilayer graphene while inset (b) the top view. The hoppings,
interaction, and gate field are also illustrated in the inset
(a).} \label{Fig:one}
\end{figure}

The ionic two-layer Hubbard model used to describe the gated bilayer graphene is given by $H=H_0+H_{int}+H_{ext},$ where
\begin{equation}
\begin{aligned}
&H_0=-\gamma_0\sum_{\langle i,j \rangle m,\sigma} (a^{\dag}_{m,i,\sigma} b_{m,j,\sigma}+H.c.) \\
&-\gamma_1\sum_{\langle i,j \rangle \sigma}(a^{\dag}_{1,i,\sigma} b_{2,j,\sigma}+H.c.) \\
&-\gamma_3 \sum_{\langle\langle i,j \rangle\rangle \sigma}(a^{\dag}_{2,i,\sigma} b_{1,j,\sigma}+H.c.)
\end{aligned}
\end{equation}
denotes the free tight-binding model containing both intralayer nearest neighbor hopping $\gamma_0$ as well as interlayer nearest (next nearest) neighbor hopping $\gamma_1$ ($\gamma_3$)~\cite{CastroGeim}, as illustrated in Fig.~\ref{Fig:one} (a).  Here,  $a_{m,i,\sigma}$ ($b_{m,i,\sigma}$) is the annihilation operator of an electron with spin $\sigma$ at site i in sublattice A (B) of layer m, and $\langle i,j \rangle$ ($\langle\langle i,j \rangle\rangle$) means the summation over nearest (next nearest) neighbor sites.
\begin{equation}
H_{int}=U\sum_{m,i} (n_{m,i,\uparrow}-1/2)(n_{m,i,\downarrow}-1/2)
\end{equation}
describes the local repulsive Coulomb interaction with $n_{m,i,\sigma}=a^{\dag}_{m,i,\sigma} a_{m,i,\sigma}$ or $b^{\dag}_{m,i,\sigma} b_{m,i,\sigma}$, depending on which sublattice the site i belongs to~\cite{explainU}. The effect of the applied perpendicular electric field is parameterized by the potential difference $\Delta$ between two layers, given by
\begin{equation}
H_{ext}=\sum_{m,i,\sigma} V_m n_{m,i,\sigma}
\end{equation}
with $V_m=-(-1)^m \Delta/2$. Here, $\Delta=eEd$ where $e$ is charge of an electron, $d=0.34$~nm is the distance between two layers, and $E=E_{ext}-4 \pi e\sum_{i,\sigma}(\langle n_{2,i,\sigma}\rangle -\langle n_{1,i,\sigma}\rangle)/S$ is the screened electric field with $E_{ext}$ the external electric field, $S$ the area of each layer and $\langle n_{l,i,\sigma}\rangle$ the average value of operator $n_{l,i,\sigma}$. Throughout the paper, $\gamma_0 = t =2.7$~eV is chosen as the unit of the energy, while $\gamma_1$ and $\gamma_3$ is fixed at $0.4$~eV and $0.3$~eV, respectively~\cite{CastroGeim}. We are only interested in half-filled case which is corresponding to the charge neutral point in experiments. The Hamiltonian without the Hubbard term expressed in momentum space is given in Appendix~\ref{app1}.

\begin{figure}[htbp]
\includegraphics[width=0.48\textwidth]{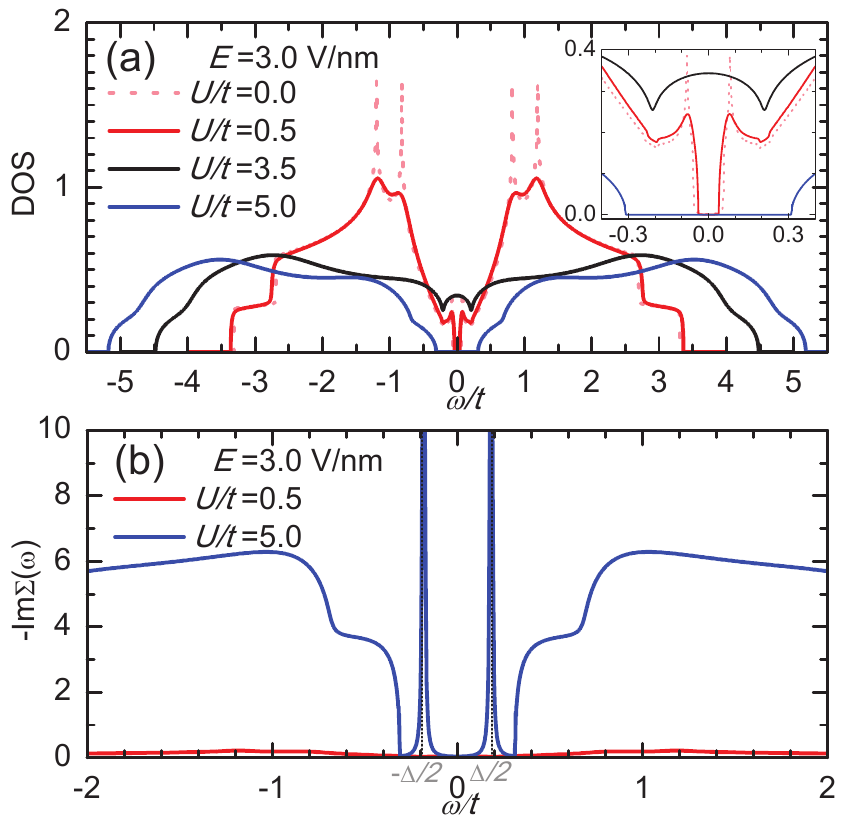}
\caption{(Color online) (a) Total density of states (DOS) for three different values of $U/t$ at a fixed value of electric field $E=3.0$~V/nm. The DOS at $U/t=0$ and $E=3.0$~V/nm is shown in dotted line. It is found that, while the system is insulating at small or large value of $U/t$, it is metallic at intermediate value of $U/t$. The inset is a blow-up of DOS around the Fermi level ($\omega=0$). (b) Imaginary parts of total self-energy for two different values of $U/t$ at $E=3.0$~V/nm corresponding to $\Delta/t=0.378$.}
\label{Fig:two}
\end{figure}

By applying the alloy analogy approach~\cite{Hubbard}, the system can be viewed as a disordered alloy where an electron with spin $\sigma$ moving on a given layer encounters either a potential of $U/2$ at a site with a spin $-\sigma$ present or $-U/2$ without, in addition to the external potential induced by the electric field. Then, the model Hamiltonian is replaced by a one-particle Hamiltonian with disorder potential which is of the form $H=H_0+H_{ext}+\sum_{m,i,\sigma} E_{m,i,\sigma} n_{m,i,\sigma}$ where $E_{m,i,\sigma}=U/2$ with probability $\langle n_{m,i,-\sigma} \rangle$ or $E_{m,i,\sigma}=-U/2$ with probability $1 - \langle n_{m,i,-\sigma} \rangle$.
The Green's function corresponding to the one-particle Hamiltonian has to be averaged over all possible disorder configurations. The averaging can not be performed exactly. To solve the alloy problem, the CPA is used~\cite{Elliot1974,Jarrell2001}. The details of the CPA method applied to the ionic two-layer Hubbard model are given in Appendix~\ref{app2}. Here, we should stress that, although above treatment itself has a few shortcomings~\cite{Gebhard}, it remains valuable as a computationally simple theory capable of capturing the Mott metal-insulator transition of many-body systems. For example, it successfully reproduces the phase diagram of an ionic Hubbard model at half filling~\cite{Hoang}.


\section{Results and discussions\label{RD}}

Now, we will show that the Coulomb repulsion does not always enhance the localization of electrons. On the contrary, it may surprisingly delocalize the electrons in the presence of gate field. Fig.~\ref{Fig:one} presents the phase diagram in the $U/t-E$ plane. In the absence of electric field, on-site Coulomb repulsion will enhance the localization of electrons as generally expected~\cite{Imada} and at a critical value of $U/t=3.5$, a single metal-to-insulator transition occurs. If without interaction, any finite gate field will impose asymmetry between layers which also leads to a gapped state~\cite{CastroGeim,McCannKoshino}. However, at a fixed value of gate voltage, tuning on the on-site Coulomb repulsion will first suppress, rather than enhance, the gapped state. As a result, an intermediate interaction-induced metallic state appears sandwiched between two insulating states.
The consecutive phase transitions from insulator to metal and then again to insulator are evident by the evolution of density of state (DOS) as a function of $U/t$. Fig.~\ref{Fig:two} (a) shows the DOS at a fixed value of gate field $E=3$~V/nm, corresponding to $\Delta/t=0.378$, for three different values of $U/t=0.5$ (insulator), $U/t=3.5$ (metal), and $U/t=5.0$ (insulator).


The nature of the two insulating states can be identified by analyzing the charge occupation number on each layer (not shown) and the total self-energy, defined as $\Sigma(\omega)=\sum_{m,i} \Sigma_{m,i}(\omega)$ where $\Sigma_{m,i}(\omega)$ denotes self-energy of site $i$ of layer $m$, in these two phases. It is found from Fig.~\ref{Fig:two} (b) that, while the imaginary part of self-energy is negligibly small at a small value of $U/t=0.5$, it becomes significantly large at a large value of $U/t=5.0$ and exhibits a divergent behavior at $\omega \simeq \pm \Delta/2$ where the two layers of bilayer graphene are located. The divergence points to the fact that the scattering rate or the effective mass of quasiparticles on each layer becomes infinite due to the strong electronic correlation. As the difference of charge occupation number between the two layers almost vanishes at $U/t=5.0$, the insulating state should be dominated by the Mott physics. On the contrary, the charge disproportionation between the layers at $U/t=0.5$ remains finite, while the imaginary part of self-energy is negligibly small, the insulator is a band insulator.

The interaction-driven metallic state at half-filling can be understood from the atomic limit. For $U<eEd$, the ground state has two electrons on each site of layer 2 and none on layer 1 due to the energy minimization, resulting in a band insulator induced by the asymmetry of charge distribution between layers. The band gap is of $eEd-U$. In the opposite condition where $U>eEd$, each site on each layer is occupied by one electron and a Mott insulator is formed with a gap $U-eEd$. Therefore, in the atomic limit, we clearly see that the interaction $U$ suppress the gap of the band insulator down to zero, but only at a single point $U=eEd$. This metallic point obtained in the atomic limit will be broaden into a metallic phase when hoppings are nonzero, as shown in our phase diagram.

\begin{figure}[htbp]
\includegraphics[width=0.48\textwidth]{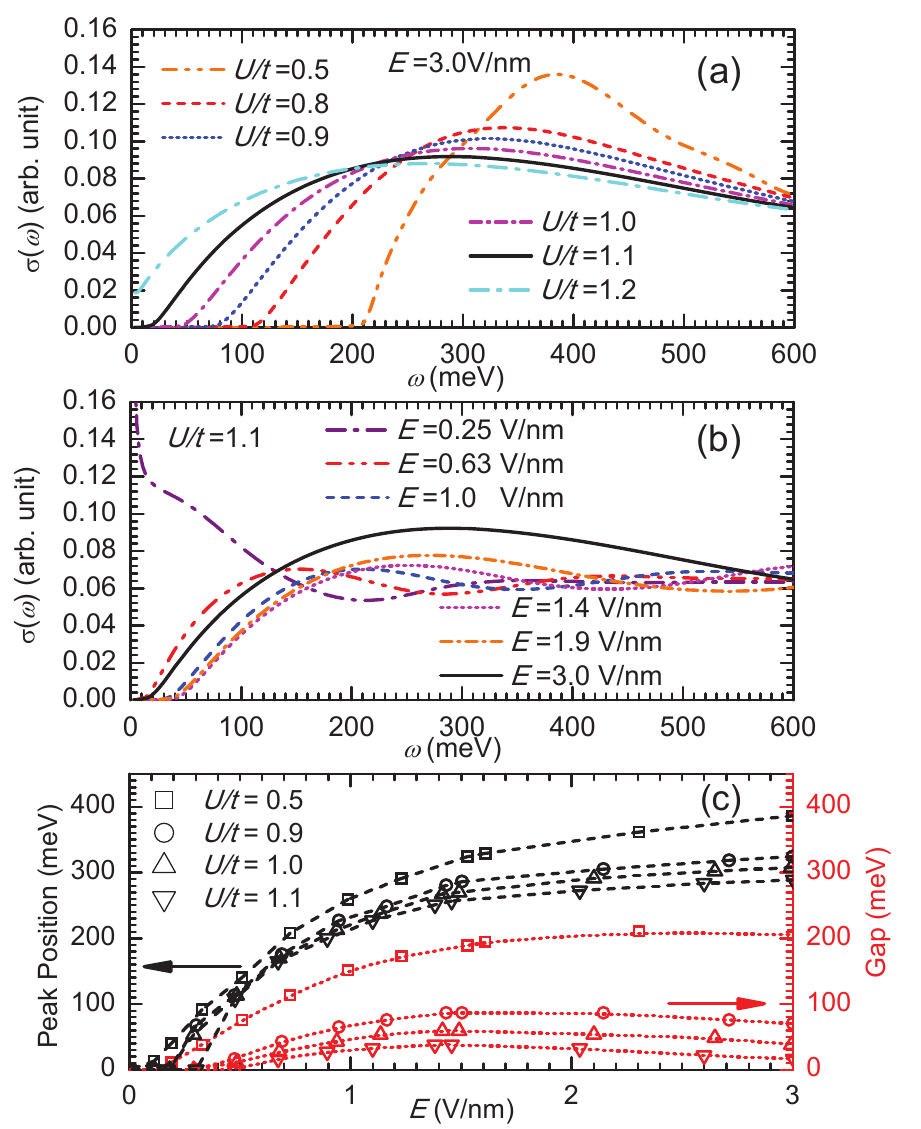}
\caption{(Color online) (a) Optical conductivities at a fixed value of $E=3.0$~V/nm for a series of $U/t$. (b) Optical conductivities at a fixed value of $U/t=1.1$ for different perpendicular electric fields $E$. (c) Peak position and the gap amplitudes in the optical conductivity as a function of electric field $E$ at four different values of $U/t=0.5, 0.9, 1.0, 1.1$.}
\label{Fig:four}
\end{figure}

Next, we will show that the long-standing experimental discrepancy regarding the gap size~\cite{McCannKoshino,Mak,Y.Zhang,Oostinga,Taychatanapat,Ulstrup} can be resolved after many-body correlation is involved. Fig.~\ref{Fig:four} (a) exhibits the optical conductivities, (see definition in Appendix~\ref{app3}), at a fixed electric field of $E=3.0$~V/nm for several values of $U/t$. Compared to the experimental results at $E=3.0$~V/nm~\cite{Y.Zhang}, the calculated optical conductivity at $U/t=1.1$ shows a broad peak below $300$~meV, as observed in the experiment. However, due to the involvement of many-body correlation, the peak position is not corresponding to the band gap. The gap is only around $16$~meV, corresponding to the edge of the optical conductivity, which is much smaller than the results derived from optical measurements ($\sim250$~meV) where the peak position is improperly taken as the band gap~\cite{Kuzmenko2009,Mak,Y.Zhang} based on an assumption that electronic correlation can be completely ignored. In fact, our result is fairly consistent with the value obtained from transport measurements ($\sim10$~meV)~\cite{Oostinga,Taychatanapat} and magnetotransport study~\cite{Varlet}, indicating that the discrepancy can be naturally ascribed to the incorrect interpretations of optical data by free tight-binding model. Compared to the previous proposals~\cite{Castro2010,Rossi2011,Santos,Kim,Park2015,Castro2008,Li2011} where midgap states induced by the lattice imperfection which were not detected in optical conductivity have to be assumed, our explanation only requires proper treatment of the inevitable correlation effect~\cite{Wehling2011,Schueler2013}.

In Fig.~\ref{Fig:four} (b), we fixed the strength of on-site Coulomb repulsion at $U/t=1.1$ and tuned the perpendicular gate field~\cite{E3Vnm}. We find that the peak position also shows monotonous gate tunability as observed in optical measurements~\cite{Y.Zhang}.
But the gap does not monotonously decrease with reduction of gate field. From $E=3.0$~V/nm to $1.4$~V/nm at $U/t=1.1$, the gap is enhanced, while starting from $E=1.4$~V/nm, the gap shrinks as gate field is reduced. As $E$ is smaller than around $0.5$~V/nm, the system becomes metallic with a Drude peak present at $\omega=0$. The metallic state may be detected in gated bilayer graphene at finite temperature region where symmetry breakings are absent. Similar situation also happens at different values of $U/t$ as shown in Fig.~\ref{Fig:four} (c) where peak positions and gaps exhibit different behaviors as a function of gate field.

\begin{figure}[htbp]
\includegraphics[width=0.48\textwidth]{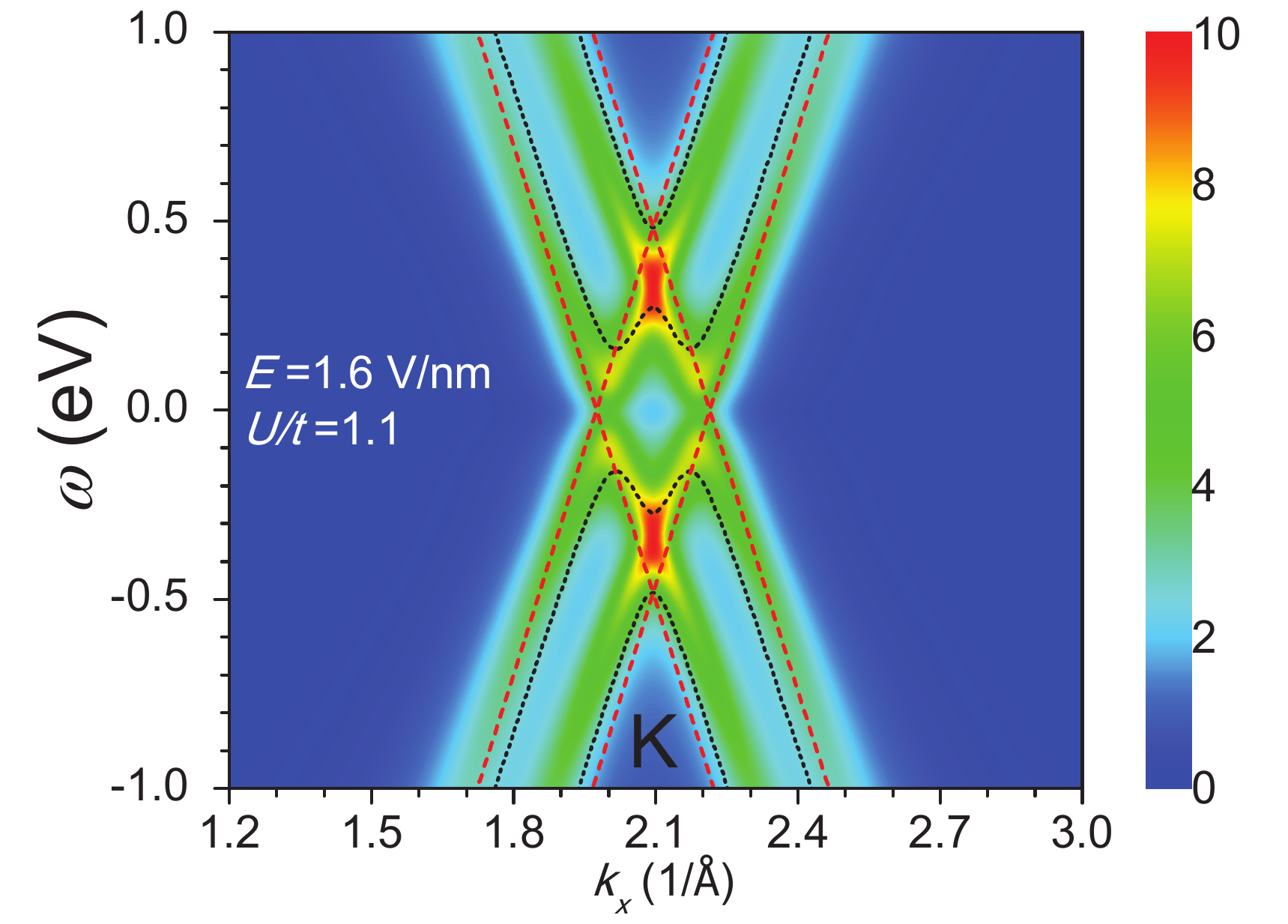}
\caption{(Color online) Spectral function along $k_x$ direction around $K$ point in momentum space. The Lorentzian broadening factor of $40$~meV is used in order to simulate the energy resolution in the angular resolved photoemission spectroscopic study. The dotted and dashed lines are the band dispersions of Bernal stacked and AA-stacking bilayer graphene, respectively, at $E=1.6$~V/nm and $U/t=0$.}
\label{Fig:five}
\end{figure}

Recently, an angular resolved photoemission spectroscopic study pointed out that the observed spectrum may indicate a coexistence of massive and massless Dirac Fermions induced by a possible imperfection of bilayer graphene~\cite{Kim}, such as a twist of the layers relative to each other which leads to a coexistence of the Bernal stacked bilayer graphene and the AA-stacking bilayer graphene where the two layers are exactly aligned~\cite{Park2015}. Here, we would like to demonstrate that the puzzling features in the spectrum can be well understood without assuming the lattice imperfection if the on-site Coulomb repulsion is included. Fig.~\ref{Fig:five} show the spectral function along $k_x$ direction around $K$ point. Again, $U/t=1.1$ is used. The electric field is set to be $1.6$~V/nm according to the experiment where the electric field is estimated from the gap at valley $K$ point~\cite{Kim}. We set a Lorentzian broadening factor of $40$~meV to simulate the energy resolution in the experiment~\cite{Kim}. It is clear from Fig.~\ref{Fig:five} that the spectral function looks like a superposition of the band structures coming from both the Bernal stacked and the AA-stacking bilayer graphene. Most importantly, prominent kinks observed experimentally which appear below and above the Fermi level, (or in other words, the neutrality point~\cite{Ohta2006,Kim}), are also present in the calculated spectrum due to the finite scattering rate induced by many-body correlation, without need to assume equal population of the AA-stacking and the Bernal stacked bilayer graphene. Such a feature is distinct from the band structure of a pure gated Bernal stacked bilayer graphene in the absence of electronic correlations (See dotted line in Fig.~\ref{Fig:five}). The vague AA-like bands between gap, derived experimentally~\cite{Kim}, may be due to the finite experimental energy resolution as well as the ignorance of suppression of gap by the Coulomb repulsion in experimental analysis.


Finally, we would like to mention that the suppression of gate-induced band gap by onsite Coulomb repulsion should always happen even in the presence of symmetry breaking. This can be first confirmed by a mean-field calculation within the Hartree-Fock approximation. Moreover, based on results of a similar model, called ionic Hubbard model, given by quantum Monte Carlo simulations~\cite{Paris}, we argue that the interaction-driven metallic state most probably persists even in the presence of magnetism.

\section{Summary\label{Summary}}

In conclusion, we use an ionic two-layer Hubbard model to study the many-body effect on the gated Bernal stacked bilayer graphene. We find that the on-site Coulomb repulsion will suppress, rather than enhance, the gate-induced gap in the bilayer graphene. By considering the on-site Coulomb repulsion, the fundamental discrepancy of gap size between transport and optical measurements is resolved and the puzzling features observed in angular resolved photoemission spectroscopic studies can be well understood, without assumption of lattice imperfection. However, in order to quantitatively describe all the experimental results, other effects like nonlocal electronic correlations, electron-phonon couplings, and disorders have to be involved. 
Our study indicate that the intrinsic and unavoidable many-body correlation should be seriously taken into account in graphene-based devices.

\section{Acknowledgments}

We thank E. V. Castro for helpful discussions. This work is supported by National Natural Science Foundation of China (Nos. 11174219 and 11474217), Program for New Century Excellent Talents in University (NCET-13-0428), and the Program for Professor of Special Appointment (Eastern Scholar) at Shanghai Institutions of Higher Learning as well as the Scientific Research Foundation for the Returned Overseas Chinese Scholars, State Education Ministry. J.-R. is also supported by Educational Commission of Anhui Province of China (No. KJ2013B059).

\appendix

\section{The Hamiltonian without the Hubbard term in momentum space\label{app1}}

By applying the Fourier transformation
\begin{equation}
c_{m,i,\sigma}=\frac{1}{\sqrt{N}}\sum_{\mathbf{k}}e^{i\mathbf{k}\cdot\mathbf{R_{i}}}c_{m,\mathbf{k},\sigma},
\end{equation}
the Hamiltonian without the Hubbard term in momentum space reads
\begin{equation}
\begin{aligned}
H_0+H_{ext}&=\sum_{k,\sigma} \Psi_{\mathbf{k},\sigma}^{\dag} \widehat{\mathcal{H}_{\sigma}}(\mathbf{k})\Psi_{\mathbf{k},\sigma},
\end{aligned}
\end{equation}
where
\begin{equation}
\widehat{\mathcal{H}_{\sigma}}(\mathbf{k})=\left[ \begin{array}{cccc}
\Delta/2                     & -t_{0}f(\mathbf{k})          & -\gamma_{1}                  & 0 \\
-t_{0}f^{*}(\mathbf{k}) & \Delta/2                          & 0 & -\gamma_{3}f(\mathbf{k})     \\
-\gamma_{1}             & 0     & -\Delta/2                         & -t_{0}f^{*}(\mathbf{k})      \\
0& -\gamma_{3}f^{*}(\mathbf{k}) & -t_{0}f(\mathbf{k})          & -\Delta/2
\end{array}
\right],
\end{equation}
and $\Psi_{\mathbf{k},\sigma}^{\dag}=(a_{1,\mathbf{k},\sigma}^{\dag},b_{1,\mathbf{k},\sigma}^{\dag},b_{2,\mathbf{k},\sigma}^{\dag},a_{2,\mathbf{k},\sigma}^{\dag})$ with
$f(\mathbf{k})=\sum\limits_{j=1}^{3}e^{-i\mathbf{k}\cdot\mathbf{\delta_{j}}}$. Here, $\mathbf{\delta_{1}}=a/2(1,\sqrt{3}), \mathbf{\delta_{2}}=a/2(1,-\sqrt{3}), \mathbf{\delta_{3}}=a(-1,0)$, and $a$ is the length of C-C bond.

\section{application of coherent potential approximation to ionic two-layer Hubbard model\label{app2}}

Hubbard~\cite{Hubbard} viewed the electron correlation problem as a disordered alloy where an electron with spin $\sigma$ moving on a given layer encounters either a potential of $U/2$ at a site with a spin $-\sigma$ present or $-U/2$ without, in addition to the external potential induced by the electric field.

So we can approximate the many-body Hamiltonian by the one-electron Hamiltonian
\begin{equation}
\label{Hamiltonian}
H=H_0+H_{ext}+\sum_{m,i,\sigma} E_{m,i,\sigma} n_{m,i,\sigma},
\end{equation}
where the disorder potential is obtained by
\begin{equation}
E_{m,i,\sigma}=\left\{
\begin{array}{lccl}
U/2& &\text{with probability}&  \langle n_{m,i,-\sigma} \rangle\\
-U/2& &\text{with probability}& 1-\langle n_{m,i,-\sigma}\rangle
\end{array}\right. ,
\label{rand.potential}
\end{equation}
here $\langle n_{m,i,\sigma} \rangle$ is the average electron occupancy per site for sublattice $i$ with spin $\sigma$ in layer $m$. The Green's function corresponding to the one-particle Hamiltonian has to be averaged over all possible disorder configurations. The averaging can not be performed exactly. To solve the alloy problem, the coherent potential approximation (CPA) is used~\cite{Elliot1974,Jarrell2001,DerwynJPCM,Derwyncpb}, where the disorder potential $E_{m,i,\sigma}$ is replaced by a local complex and energy-dependent self-energy.

Then, the Hamiltonian within CPA becomes
\begin{equation}
H_{CPA}=H_0+H_{ext}+\sum_{m,i,\sigma}\Sigma_{m,i,\sigma} n_{m,i,\sigma}.
\label{CPA Hamiltonian}
\end{equation}
And corresponding CPA Hamiltonian in momentum space reads
\begin{equation}
H_{CPA}=\sum_{k,\sigma} \Psi_{\mathbf{k},\sigma}^{\dag} \widehat{\mathcal{H^{CPA}}_{\sigma}}(\mathbf{k})\Psi_{\mathbf{k},\sigma},
\end{equation}
where
\begin{equation}
\begin{tiny}
\widehat{\mathcal{H^{CPA}}_{\sigma}}(\mathbf{k})=\left[ \begin{array}{cccc}
\frac{\Delta}{2}+\Sigma_{1A\sigma} & -t_{0}f(\mathbf{k})         &  -\gamma_{1}                  & 0 \\
-t_{0}f^{*}(\mathbf{k})   & \frac{\Delta}{2}+\Sigma_{1B\sigma}   & 0 & -\gamma_{3}f(\mathbf{k})     \\
-\gamma_{1}               & 0   &  -\frac{\Delta}{2}+\Sigma_{2B\sigma}   & -t_{0}f^{*}(\mathbf{k})      \\
0 & -\gamma_{3}f^{*}(\mathbf{k})&  -t_{0}f(\mathbf{k})          & -\frac{\Delta}{2}+\Sigma_{2A\sigma}
\end{array}
\right].
\label{CPA Hamil.moment.}
\end{tiny}
\end{equation}
The CPA average Green's function can be written in matrix form
\begin{equation}
\begin{tiny}
\bar{G}(\mathbf{k},\omega)=\left[ \begin{array}{cccc}
\omega-\frac{\Delta}{2}-\Sigma_{1A} & t_{0}f(\mathbf{k})               & \gamma_{1}                       & 0       \\
t_{0}f^{*}(\mathbf{k})           & \omega-\frac{\Delta}{2}-\Sigma_{1B} & 0     & \gamma_{3}f(\mathbf{k})           \\
\gamma_{1}                       & 0         & \omega+\frac{\Delta}{2}-\Sigma_{2B} & t_{0}f^{*}(\mathbf{k})            \\
0          & \gamma_{3}f^{*}(\mathbf{k})      & t_{0}f(\mathbf{k})               & \omega+\frac{\Delta}{2}-\Sigma_{2A}
\end{array}
\right]^{-1},
\end{tiny}
\end{equation}
where all spin indices have been omitted as we are interested in the paramagnetic phase. In real space, we have
\begin{equation}
\bar{G}_{mi,mi}(\omega)=\frac{1}{\Omega_{BZ}}\int_{\Omega_{BZ}}d\mathbf{k} \bar{G}_{mi,mi}(\mathbf{k},\omega),
\label{real Green}
\end{equation}
where the integral is over the first Brillouin zone of the sublattice. Then a cavity Green's function $\mathcal
{G}_{mi}(\omega)$ can be obtained through the Dyson equation
\begin{equation}
\mathcal {G}_{mi}^{-1}(\omega)=\bar{G}_{mi,mi}^{-1}(\omega)+\Sigma_{mi}(\omega)
\end{equation}
for sublattice ($i=A,B$) in each layer ($m=1,2$), which describes a medium with self-energy at a chosen site removed. The cavity can now be filled by a real "impurity" with disorder potential, resulting in an impurity Green's function
\begin{equation}
G_{mi}^{\gamma}(\omega)=[\mathcal {G}_{mi}^{-1}(\omega)-E_{m,i}^{\gamma}]^{-1}
\end{equation}
with impurity configurations of
$E_{m,i}^{\gamma}=\begin{cases}U/2&\gamma=+\\
-U/2&\gamma=-\end{cases}$ as defined by Eq.(\ref{rand.potential}). The CPA requires
\begin{equation}
\langle G_{mi}^{\gamma}(\omega)\rangle =\bar{G}_{mi,mi}(\omega),
\label{ave.Green}
\end{equation}
where the average is taken over the impurity configuration probabilities defined by Eq.(\ref{rand.potential}).

Equation(\ref{real Green}) and (\ref{ave.Green}) need to be solved self-consistently. Since electrons of bilayer graphene under the electric field prefer to be on the layer 2 for $\Delta>0$ and the condition that $\sum\limits_{m,i}\langle n_{m,i}\rangle=4$ at half-filling must be satisfied, where
\begin{equation}
\langle n_{m,i}\rangle=-\frac{1}{\pi}\int_{-\infty}^{0}Im\bar{G}_{m,i}d\omega,
\end{equation}
the resulting integrated DOS for each site should be consistent with the average occupation number probabilities used in Eq.(\ref{ave.Green}), so an extra loop of self-consistency should be added.

\section{calculation of optical conductivity\label{app3}}

The optical conductivity is defined as~\cite{Rozenberg,Pruschke}

\begin{equation}
\begin{aligned}
 \sigma(\omega)=\frac{2e^2t^2a^2}{v\hbar^2}\int_{-\infty}^{+\infty}d\epsilon D(\epsilon)\int_{-\infty}^{+\infty}\frac{d\omega^{'}}{2\pi}\rho(\epsilon,\omega^{'})\\
\rho(\epsilon,\omega^{'}+\omega)\frac{f(\omega^{'})-f(\omega^{'}+\omega)}{\omega},
\end{aligned}
\end{equation}
where $e$ is the electron charge, $a$ is lattice constant, $v$ is the volume of primitive cell, $t$ is nearest neighbor hopping energy, $D(\epsilon)$ is free particle density of state, and
\begin{equation}
\begin{aligned}
\rho(\epsilon,\omega)=-\frac{1}{\pi}ImG(\epsilon,\omega)=-\frac{1}{\pi}Im\frac{1}{\omega+i\eta+\mu-\epsilon-\Sigma(\omega)},
\end{aligned}
\end{equation}

\begin{equation}
\begin{aligned}
f(\omega)&=\frac{1}{e^{\beta\omega}+1},
\end{aligned}
\end{equation}
with $\beta=\frac{1}{kT}$. For $T=0K$, the optical conductivity can be abbreviated as
\begin{equation}
\begin{aligned}
 \sigma(\omega)&=\frac{2e^2t^2a^2}{v\hbar^2}\int_{-\infty}^{+\infty}d\epsilon D(\epsilon)\int_{-\omega}^{0}\frac{d\omega^{'}}{2\pi\omega}\rho(\epsilon,\omega^{'})\rho(\epsilon,\omega^{'}+\omega)
\end{aligned}
\end{equation}


\begin{thebibliography}{99}
\bibitem{explainU} Please note $U$ denotes strength of the onsite Coulomb interaction between electrons. While it is insusceptible to the gated electric field, the relative strength of $U$ with respect to hoppings can be tuned by stretching bilayer graphene within a-b plane.
\bibitem{E3Vnm} The reason why we only compare our result with experimental one for $E=3$~V/nm is that, in the experimental paper (Fig. 2 of Ref.~\onlinecite{Y.Zhang}), only the curve for $E=3$~V/nm clearly shows a prominent peak which can be attributed to the transition of electrons from the valence band to the conduction band, in addition to a very sharp spectral feature which is attributed to the zone center G-mode phonon in graphene. Other curves either do not show any definite peak or only present one broad peak with complicated structure which may be due to the mixture of the peaks contributed from phonons and electrons.
\end{thebibliography}
\end{document}